\documentclass[]{interact}

\usepackage[english]{babel}

\usepackage{ragged2e} 
\usepackage{amsmath,amsthm,amsfonts,amssymb,amscd,mathrsfs}
\usepackage{graphicx}
\usepackage{bm,xstring}
\usepackage{xcolor}
\usepackage{placeins}
\usepackage[colorlinks=true, allcolors=blue]{hyperref}
\usepackage{booktabs}
\usepackage{tabularx, ltablex}
\usepackage{lscape}
\usepackage{multirow}
\usepackage{dsfont}
\newcolumntype{s}{>{\setbox0=\hbox\bgroup}c<{\egroup}@{}} 
\usepackage{cuted}
\usepackage[T1]{fontenc}
\usepackage{orcidlink}
\usepackage{subcaption}
\usepackage[utf8]{inputenc}

\usepackage[natbibapa,doi]{apacite}
\usepackage{bibentry}
\usepackage{caption}
\usepackage{subcaption}
\usepackage{etoolbox}
\usepackage{environ}
\usepackage{keyval}
\newtoggle{bibdoi}
\newtoggle{biburl}
\newtoggle{publisher}
\newtoggle{address}
\makeatletter
\undef{\APACrefURL}
\undef{\endAPACrefURL}
\undef{\APACrefDOI}
\undef{\endAPACrefDOI}
\long\def\collect@url#1{\global\def\bib@url{#1}}
\long\def\collect@doi#1{\global\def\bib@doi{#1}}
\newenvironment{APACrefURL}{\global\toggletrue{biburl}\Collect@Body\collect@url}{\unskip\unskip}
\newenvironment{APACrefDOI}{\global\toggletrue{bibdoi}\Collect@Body\collect@doi}{}
\AtBeginEnvironment{thebibliography}{
\pretocmd{\PrintBackRefs}{%
\iftoggle{bibdoi}
{
\iftoggle{biburl}{\unskip\unskip doi:\ignorespaces\bib@doi}{doi:\ignorespaces\bib@doi}} 
{\iftoggle{biburl}{Retrieved from \bib@url}{}}
\togglefalse{bibdoi}\togglefalse{biburl}%
}{}{}
}

\newcommand*{\doi}[1]{\href{https://doi.org/#1}{\detokenize{#1}}}

\usepackage{credits}

\credit{NF}  {0,1,1,0,1,1,1,0,1,0,0,1,1,1}
\credit{JS}  {1,1,1,0,1,1,1,0,0,1,1,1,1,1}
\credit{KG}  {0,1,1,0,1,1,0,0,1,0,1,1,1,0}
\credit{MP}  {0,0,0,1,0,1,0,0,0,1,0,0,0,1}
\credit{PD}  {1,0,0,1,0,1,0,0,0,1,0,0,0,1}

\AtBeginDocument{}
\begin{document}

\title{Adapting tree-based multiple imputation methods for multi-level data? A simulation study}

\author{Nico Föge\textsuperscript{a,*}\thanks{Communicating author: Nico Föge, TU Dortmund University, Email: \texttt{nico.foege@ovgu.de}.}, Jakob Schwerter\textsuperscript{b,c,*}, Ketevan Gurtskaia\textsuperscript{b}, Markus Pauly\textsuperscript{b,d}, and Philipp Doebler\textsuperscript{b} \\
\textsuperscript{a}Department of Mathematics, Otto-von-Guericke Universität Magdeburg, Germany \\
\textsuperscript{b}Department of Statistics, TU Dortmund University, Dortmund, Germany \\
\textsuperscript{c}Center for Research on Education and School Development, TU Dortmund University, Dortmund, Germany \\
\textsuperscript{d}Research Center for Trustworthy Data Science and Security, UA Ruhr, Germany \\
\textsuperscript{*}Shared first authorship}

\maketitle
\begin{abstract}
When data have a hierarchical structure, such as students nested within classrooms, ignoring dependencies between observations can compromise the validity of imputation procedures. Standard (tree-based) imputation methods implicitly assume independence between observations, limiting their applicability in multilevel data settings. Although Multivariate Imputation by Chained Equations (MICE) is widely used for hierarchical data, it has limitations, including sensitivity to model specification and computational complexity. Alternative tree-based approaches have shown promise for individual-level data, but remain largely unexplored for hierarchical contexts. In this simulation study, we systematically evaluate the performance of novel tree-based methods -- Chained Random Forests (missRanger) and Extreme Gradient Boosting (mixgb) -- explicitly adapted for multi-level data by incorporating dummy variables indicating cluster membership. We compare these tree-based methods and their adapted versions with traditional MICE imputation in terms of coefficient estimation bias, type I error rates and statistical power, under different cluster sizes (25 and 50), missingness mechanisms (MCAR, MAR) and missingness rates (10\%, 30\%, 50\%), using both random intercept and random slope data generation models. The results show that MICE provides robust and accurate inference for level 2 variables, especially at low missingness rates (10\%). However, the adapted boosting approach (mixgb with cluster dummies) consistently outperforms other methods for Level-1 variables at higher missingness rates (30\%, 50\%). For level 2 variables, while MICE retains better power at moderate missingness (30\%), adapted boosting becomes superior at high missingness (50\%), regardless of the missingness mechanism or cluster size. These findings highlight the potential of appropriately adapted tree-based imputation methods as effective alternatives to conventional MICE in multilevel data analyses.
\end{abstract}

\begin{keywords}
bias; hierarchical data; 
MICE; missRanger; mixgb; Power; Type I error
\end{keywords}

\newpage

\section{Introduction}
\label{introduction}
 A prominent and widely used technique for dealing with missing data is multiple imputation (MI), which involves creating multiple reasonable values for each missing value, so that multiple complete data sets are generated \citep{littlerubin}. Each of the complete data sets is used separately to perform statistical analysis using standard statistical techniques. The results of the analyses are then combined using methods that account for the variability of the imputed values, resulting in more reliable estimates \citep{rubin}. MI is particularly useful when a large amount of data is missing and the missingness mechanism is missing at random or missing completely at random \citep{schafer, littlerubin}. 

Ensuring that imputation models preserve the underlying relationships in the data and accounting for the missing data mechanism is crucial \citep{micer}. 
This is a particular challenge in complex hierarchical data sets with multiple levels. Examples cover clustering at individual and higher-level units like classes or schools. Here, maintaining the hierarchical structure is  often overlooked when imputing missing values which may cause biased analyses \citep{vanBuuren2011multiple}.

Although fully conditional specification (FCS) or multivariate imputation by chained equations (MICE) is a prevalent approach in the social sciences, it presents several limitations, including its considerable complexity due to the challenges of the specification of imputation model and computational intensity \citep{mice}. Since MICE heavily relies on model specifications, it can lead to issues like overfitting and convergence errors, especially when dealing with multicollinearity and other instability problems \citep{van1999flexible}.

Exploring alternatives to MICE, such as non-parametric tree-based methods can enhance the robustness and reliability of statistical analysis in empirical studies.
While tree-based methods have been increasingly used for
imputatuion in single-level data, their application in the context of multilevel data remains largely unexplored.

Therefore,  this simulation study addresses the following \textit{research question}: Do tree-based imputation methods exhibit similar performance in terms of bias, type I error, and power compared to the standard level-2 imputation method? The specific tree-based methods are chained random forests \citep{missforest,RFTang} and extreme gradient boosting  with trees as base learners \citep{mixgb}. To account for the multilevel data structure, we use dummy variables for each cluster,  adapting tree-based imputation methods 
in the same way \citet{LudtkeRG2017} did for non-tree-based methods.

Although many different methods 
MI have been proposed and investigated for this setting -- including multilevel MI with joint modeling, multilevel MI with fully conditional specifications, multilevel substantive-model-compatible MI with sequential modeling, or model-based treatment with Bayesian estimation, as  recently reviewed by \cite{GrundLR2024} -- tree-based methods have not been evaluated for multilevel data.

The remainder of the paper is organized as follows: In the next section, we briefly review multilevel data and existing (parametric) MI approaches for multilevel data structures. Next, the simulation study is detailed in Section 3, starting with a review of the imputation approaches and an explanation of the simulated missingness mechanisms. We present simulation results in Section 4 also with the help of bar plots for the relative performance, and we close in Section 5 with remarks on the current study and potential future work.

\section{Multilevel data} \label{ch:multilevel}

Multilevel data structures are common in social and behavioral sciences research  \citep{Handbook}. This is often seen in, e.g., educational research where students (Level 1) are nested within classes, schools, or regions (Level 2). 
The presence of higher-level variables can significantly influence the outcome variable. This calls for robust analysis methods that account for the complexity introduced by these hierarchical structures \citep{ModelingMDS, GrundLR2024}, since simply ignoring the dependencies or aggregating everything to a single level can be deceptive \citep{aitkinandLongford, GrundLR2024}.

To maintain statistical integrity, it is crucial to employ appropriate imputation methods tailored for multilevel data. These methods should consider both within-cluster and between-cluster variability for a more accurate representation of the underlying data \citep{vanBuuren2011multiple,j46, GrundLR2024}. Ignoring the nested structure of data through aggregation or disaggregation to a single level is suboptimal and can result in misleading conclusions \citep{aitkinandLongford}. To address this,  multilevel modeling techniques have been developed to address the variance components properly at each level. In particular,  hierarchical linear models, including the random intercept model as well as the random intercept and random slope model, provide a sophisticated statistical framework for analyzing hierarchical data structures \citep{hox, Handbook}.

\subsection{Existing imputation approaches for multilevel data}

For a valid analysis, the 
imputation model must account 
for the dependency between observations inherent in the multilevel structure. Otherwise, inferences might be biased even when the statistical methods are appropriate \citep{Handbook,j46}. 
The bias of random-effect variances as well as global fixed-effects confidence intervals depends
on the cluster size, the relation of within- and between-cluster variance and the missing data mechanism \citep{speidel2018biases}. 
A study from \cite{endetal16} compared two imputation frameworks for multilevel data: joint modeling (JM) and chained equation imputation (MICE with a two-level normal model \verb|2l.norm|). The joint model turned out to be better for analysis postulating distinct within- and between-cluster relations while chained equations imputation turned out to be superior in random slope analysis \citep{endetal16}. With JM, imputations are created according to a joint model for all variables simultaneously 
drawn from the fitted distribution 
\citep[van Buuren in][pp. 173-196]{Handbook}. 

Next to JM, another standard MI procedures for multivariate multilevel data is a \textit{fully conditional specification} of MI ({FCS}). Imputations with {FCS} are constructed on a variable-by-variable basis. For each variable with missing parts, an imputation model is specified and imputations are iteratively generated by cycling through the variables and imputation model \citep{buuren}.

For clustered data both of these methods are effective in the broad context of \textit{random intercept} models, even with variables at higher levels, as simulation results by  \citet{grundluedtkesimulationsandreccomendations} indicate.  However,  for \textit{random slopes} models, FCS appears to be more flexible than JM, but still has some limitations and is not all that reliable when data in an explanatory variable is missing \citep{grundluedtkesimulationsandreccomendations}. Similar results were shown by \citet{endetal16} for \textit{random intercepts}. 
Additionally, \citet{grundluedtkeLEVEL2} show that both approaches (chained equations and joint modeling) provide useful tools for dealing with missing data at Level 2 in most applications in practice (especially for balanced data). 
 
 One advantage of choosing FCS over JM is that FCS grants more flexibility in creating multilevel models by splitting a $k$-dimensional problem into $k$ one-dimensional problems. That is, for each of the variables with missing data, a regression model with a univariate outcome conditional on the other $k-1$ variables is constructed. Furthermore, it is easier to avoid logical inconsistencies in the imputed data and incorporate methods to preserve unique features in the data \citep{BuurenFCS}, e.g., temporal dependency can be taken into account in longitudinal data.  Multivariate imputation by chained equations (MICE), often used synonymously to FCS, is a specific implementation of the broader FCS framework. MICE uses the conditional imputation approach from FCS and chains the variables by imputing iteratively one variable at a time based on the others \citep{micer}.

\subsection{Multiple imputation methods}

In this simulation study, we implemented three main imputation methods: 
As baseline we use 
multivariate imputation by chained equations (MICE) via
the \texttt{mice} function from the \texttt{R} package of the same name \citep{mice}.
For Level 1 variables the Level 1 normal model (\verb|2l.norm|) is used and for Level 2 variables the Level 2 class predictive mean matching (\verb|2lonly.pmm|).
Additionally, we include recent tree-based imputation methods:  a fast implementation of imputation via random forests from the \verb|missRanger| package \citep{missRanger}, and multiple imputation by XGBoost implemented in the \verb|mixgb| package \citep{mixgb}. For MICE the imputation methods and the prediction matrix for each variable are carefully adjusted to take into account the multilevel data structure.  Two factors were varied for \verb|missRanger|: 
We run the \verb|missRanger| algorithm both with and without predictive mean matching (using 5 donors)\footnote{The default value for \texttt{missRanger} is 0, and the default value in MICE is 5. Not using PMM for \texttt{missRanger} showed inflated type I errors in another context \citep{RamosajAP2020}.}
and the variant (standard or an adapted implementation of \verb|missRanger|). The adapted variant respects the multilevel structure of the data by including dummy variables for the respective cluster, while the standard implementation does not.
In line with \cite{drechsler2015multiple}, who shows that multilevel models at the imputation stage can provide superior inference—especially for random effects—we acknowledge that running such models can be computationally demanding and that researchers often resort to simpler, dummy-based approaches. Indeed, representing cluster membership via dummies biases random‐effects estimates more than fixed‐effects estimates, yet remains popular because it is straightforward to implement with many conventional imputation tools. As we pioneer the adaption of tree-based methods to hierarchical data, we follow Drechsler’s rationale and incorporate dummy variables for clusters in our adjusted version of \texttt{missRanger}. While this does not capture the complete random‐effects structure, it delivers a manageable, first‐tier solution for incorporating Level 2 distinctions without the overhead of fully specified multilevel modeling.
 Similarly,  \verb|mixgb| was adapted with additional dummy variables. We also used 5 imputations\footnote{To choose a number of imputations a mini-simulation on randomly selected design was run with $5$ and $20$ imputations with MICE and \texttt{missRanger} (standard and with dummies with three donors). The results suggested very little to no differences between the number of imputations in terms of Coefficient estimation bias and $H_0$ rejection rates, but imputing $20$ times takes approximately four times longer than performing $5$ imputations. Consequently, the whole simulation is run with $5$ imputations for each method, making the simulation considerably faster.}. Details of each method are given in the sequel.

\subsubsection{Multivariate imputation by chained equations}
\textit{Multivariate Imputation by Chained Equations} (\textbf{MICE}) is a flexible and efficient imputation method that can treat missingness in a wide range of data types and analysis models. Its applications can be found in diverse research fields, including medicine, epidemiology, psychology, management, politics, and sociology \citep{mice}. 
At its core, MICE operates through Gibbs Sampling, iteratively drawing missing values from univariate conditional distributions given the other observed features. By repeatedly updating these imputations over several iterations, the method approximates the underlying conditional distribution, simulating the burn-in phase of a Markov chain. Once the process stabilizes, the multiple imputations are generated. The idea behind is that the variability between imputations arises from independent draws from this stable conditional distribution \citep{mice,kistner2024enhancing}.

Imputing at Level 2 requires additional considerations. One possibility is to have a separate model for Level 2 variables, for example, a regression model that includes other Level 2 variables as well as cluster-level components (e.g, mean, median) of the Level 1 variables.  As described by \citet{grundluedtkeLEVEL2}, imputations at Level 2  are generated based on a probability distribution that considers multiple factors. The missing values for a variable at Level 2 within a cluster depend on the aggregated information from Level 1 (such as group means of relevant variables), the observed values of the other variables within the cluster, and specific model parameters.

As mentioned, \verb|mice| is an R package for chained equations imputations using a function of the same name. To impute Level 1 missing values \verb|2l.norm|\footnote{2l indicates the data structure which is multi-level, it does not mean the calculation happens at a higher level, but it considers that there are variables at another level} is used, which uses univariate missing data imputation with a two-level normal model. For Level 2 variables,  a level-2 class predictive mean matching (\verb|2lonly.pmm|) is applied. 
Although \cite{endetal16} used \verb|2lonly.norm|, we use \verb|2lonly.pmm| because \verb|2lonly.norm| follows the normality assumption, which does not hold in our case because we transformed some variables to aggregate them at Level 2. \verb|2lonly.pmm| is a semi-parametric method, which is why it works in a wider range of cases.

\subsubsection{Chained Random Forest imputation}
The idea of the Random Forest (RF) is to combine predictions from multiple decision trees to improve accuracy and reduce overfitting in both classification and regression tasks. In a Random Forest, each tree independently makes a prediction, and the final result is based on the majority vote (in classification) or the average prediction (in regression) of all the trees. RF can handle mixed types of data, is capable of addressing interactions and nonlinearity, and is robust to overfitting \citep{Breiman}.
 Due to its flexibility, RF can also be adapted to imputation tasks \citep{missforest,golino2016random} transferring useful properties, such as the lack of assumptions about normality or homoscedasticity. 
 
 As for MICE, the chained Random Forest imputation also starts with an initial guess, that
 replaces all missing values in the data.
 The initial guess can be made by mean (default for metric data), median, mode (default for categorical data), or any other univariate imputation method.
 At this point, every variable in the dataset is complete, although the imputations are just placeholders. Next, the columns in the dataset are ranked according to the proportion of missing data, starting with the column containing the fewest missing values.
 For each column, the algorithm builds a predictive model using Random Forests, where the observed (non-missing) values in that column serve as the outcome variable, and the values from other columns for the same rows are used as predictors. Once predictions are generated, the initial guesses are replaced by these predictions.
 After cycling through all variables with missing values (one full iteration), the algorithm repeats the process $\texttt{maxiter=10}$ times
 \citep{missforest}. During each iteration, the imputed values from the previous step are updated using the predictions from the Random Forest models trained in the current iteration.
 As the iterations proceed, the imputations are refined, as the predictions are based on increasingly accurate imputed values for the other variables.
  The algorithm stops after a pre-defined number of iterations. In the present paper, we determine this number with the help of a small simulation.

A faster and extended version of the algorithm is given by  
the \verb|missRanger|  package of \cite{missRanger}. It uses the run-time efficient \textit{ranger} implementation of RF 
and additionally extends \verb|missForest| by offering the option of \textit{predictive mean matching} \citep[PMM;][]{missRanger}. PMM enhances Random Forest imputation by ensuring imputed values are more realistic. Instead of directly using predicted values from the Random Forest, PMM matches each missing entry to one of the $k$ closest observed values, the so called donors. The actual observed value from this match is then used for imputation, maintaining the variability of the original data.
PMM is particularly useful when it is important to faithfully reflect the underlying data distribution and impute values that are plausible within the observed range \citep{missRanger}. 
The number of donors is denoted by $k$. The default value is $k=5$ which we also use in our study.

\subsubsection{MI through extreme gradient boosting (XGBoost)}

Extreme Gradient Boosting (XGBoost) is a machine learning algorithm that belongs to the family of gradient boosting methods \citep{ChenG2016}. Instead of simply averaging multiple decision trees as in Random Forests, XGBoost uses gradient boosting to combine multiple regression trees. It also employs regularization and shrinkage techniques. Typically, the depth of the regression trees used in sequential boosting are not very deep. One advantage of XGBoost over Random Forest is that XGBoost selects only one of a set of highly correlated features, because in sequential boosting, features are added to the model incrementally. If a feature is selected early, other highly correlated features can only improve the prediction if they provide new information.

XGBoost uses gradient descent optimization techniques to iteratively minimize the loss function, making it highly efficient and effective in finding the optimal solution. L1 and L2 regularization prevent overfitting and improve generalization. Tree pruning removes unnecessary branches and reduces model complexity, further enhancing XGBoost's predictive performance. Lastly, XGBoost can take advantage of parallel processing capabilities, making it suitable for large datasets and reducing training time. Overall, XGBoost is known for its ability to handle complex datasets and provide accurate predictions. Its popularity is evident in various domains, including Kaggle competitions and real-world applications \citep{ChenG2016}.

The \verb|mixgb| \verb|R| package uses XGBoost to implement missing value imputation in a scalable and efficient manner \citep{DengY2023}. It addresses the challenge of missing data in large datasets with complex structures. \verb|mixgb| imputes missing values in the order of variables with fewer missing values. This strategic approach aims to prioritize variables with more available information during the imputation process. Initial values for imputation are filled with random values drawn from the observed data to kickstart the imputation process \citep{suh2023comparison}. \verb|mixgb| provides a versatile approach to missing data imputation, leveraging XGBoost, subsampling, and PMM to enhance imputation accuracy, especially for continuous data. The default imputation in \verb|mixgb| is non-iterative but the package allows users to set the number of iterations for imputation \citep{DengY2023}. For this simulation we used 5 iterations to be in line with the other imputation methods.

\subsubsection{Adapting tree-based imputation methods with cluster membership dummy variable}
By default, both, \texttt{missRanger} and \texttt{mixgb}, are unaware of the multilevel data structure. We study variants of the two procedures that add dummy variables for cluster membership \citep{LudtkeRG2017}. If there are $J$ clusters, $J$ dummy variables are added, and the $j$th variable is equal to $1$ if the data in row $i$ is from cluster $j$ and $0$ otherwise. 
In contrast to the regression methods studied in \cite{LudtkeRG2017}, no reference group is necessary for the tree methods. Adding the dummy variables is possible in all standard software packages. The idea is that cluster-level effects could now influence the imputations, as the models can generate cluster-specific predictions.
After imputation, level-2 variables are aggregated to ensure consistency within clusters. Specifically, the mean value is computed for each cluster, and this aggregated value is assigned to all observations within the respective cluster. The repeated assignment ensures that all observations in the same cluster share the same level-2 information, preserving the hierarchical data structure.

If regression trees have high enough tree depth, the flexibility gained by the dummy variables is potentially higher compared to a procedure that would include some kind of random intercepts, since some interactions with the dummy variables are possible,  though not all kinds of interactions \citep{wright2016little}. Potentially, more dummy variables are added than there are variables originally, so the dummy variable procedure relies on imputation methods that have some form of regularization. Tree-based ensemble methods like RF and XGBoost do have this property by design \citep{Breiman, ChenG2016}.

\section{Simulation study design}

The use of tree-based methods  grows in  empirical research, especially with \verb|missRanger| \citep{Sajeev2022, Sommer2023, SchwerterBDM2023metropolitan, kistner2024enhancing}. However, it is still unclear how reliable the statistical inference is for data that has been imputed with tree-based methods. 
While a recent simulation study by \cite{Seminarreport} investigated the performance of tree-based methods in longitudinal data, we focus on their performance on hierarchical data, especially when there is missingness at a higher level. 
In particular, we aim to find the strengths and limitations of tree-based imputation methods concerning type I error rates, statistical power, and coefficient bias, compared to the widely-used MICE approach. To this end, we simulated multilevel data with data missing at random, completely at random and not at random. For the computations \texttt{R} Version 4.4.0 was used \citep{R}.

\subsection{Simulated data}

For the simulation setup, four general factors are varied, resulting in a total of 24 simulation designs. The varying factors are: \textbf{number of clusters, data generation model, missing rate, and missing mechanism}. We consider two numbers of clusters: 25 and 50. These choices reflect typical applied scenarios, with 25 clusters representing a small-sample setting (<30 clusters) and 50 clusters approximating asymptotic conditions \cite{CameronM2015}. This distinction matters practically, as inference methods perform differently depending on whether the number of clusters is small or large.
This results in two cluster sizes of 40 and 20 for a balanced design with a sample size of $\boldsymbol{N} = 1000$. This variation in the number of clusters allows us to examine the effect of cluster size on the performance of the imputation methods

\subsection{Data generation}
To assess the impact of these factors, we simulate  data in a controlled manner. Specifically, we generate data by using the \verb|fungible| package in \verb|R| \citep{fungible}.
The \verb|monte| function generates clustered data with predefined characteristics. For this simulation study, intra-cluster correlations and indicator validities (cluster separations for each variable) are randomly constructed. Six variables are Level 1 (individual level) variables and six variables are aggregated to Level 2. All covariates are continuous numerical variables. Finally, two data generation models are considered for the outcome variable: \textit{random intercept} and \textit{random intercept and random slope} (hereafter denoted as \textit{random slope} model). Based on the model, the output variable is entered with 

\begin{align}
Y_{ij}^{o} &= 0.3 \nonumber\\
&\quad + \delta_{0j} + 0.5 \times X_{ij1}+1 \times X_{ij2}+1.5 \times X_{ij3}+0 \times X_{ij4} 
+0 \times X_{ij5}+0 \times X_{ij6}\nonumber\\
&\quad +0.5 \times L_{ij1} +1 \times L_{ij2}+1.5 \times L_{ij3}+
+0 \times L_{ij4} +0 \times L_{ij5} +0 \times L_{ij6}+\epsilon_{ij} \nonumber\\
\intertext{for the random intercept model and}
Y_{ij}^{o} &= 0.3 \nonumber \\
&\quad + \delta_{0j} + (0.5+\delta_{1j}) \times X_{ij1}+(1+\delta_{2j}) \times X_{ij2}+1.5 \times X_{ij3}+0 \times X_{ij4} \nonumber\\
&\quad+0 \times X_{ij5}+0 \times X_{ij6}\nonumber\\
&\quad +0.5 \times L_{ij1} +1 \times L_{ij2}+1.5 \times L_{ij3}
+0 \times L_{ij4} +0 \times L_{ij5} +0 \times L_{ij6}+\epsilon_{ij} \nonumber
\end{align}
in case of the random slope model, where $\delta_{0j}, \delta_{1j}, \delta_{2j}, \epsilon_{ij} \sim \mathcal{N}(0,1)$ are randomly generated.
The six noise features $Y_{ij4}, Y_{ij5}, Y_{ij6}, L_{ij4}, L_{ij5}, L_{ij6}$
are uncorrelated with all other features. They are included to enable evaluation of the type one error and to assess the
ability of the imputation methods to distinguish  between important and unimportant variables.

\subsubsection{Missingness mechanism}

Missingness in the generated data is induced at five levels from moderately low to relatively high ($10\%$, $30\%$ and $50\%$). This range of missingness levels reflects real-world scenarios. Two missingness mechanisms (MAR, MCAR) are considered. The introduction of missing data is based on the selected missingness mechanism and the specified missing rate. For MCAR, a simple function is implemented in \verb|R| that randomly sets each data point to \textbf{NA} with probability equal to the missingness level for each variable. To introduce missingness according to the MAR mechanism, the algorithm described by \cite{Thurow2021,thurow2021imputing} is used, with slight adjustments for purely numerical data. The algorithm operates as follows: starting with one variable it generates missing values under MCAR with the overall missing rate. The missingness in all other variables depends only on the remaining (observed) data in the first variable to exhibit MAR missingness. After introducing missingness in the first variable, it is converted to a categorical variable by grouping values into intervals. For each subsequent variable i.i.d. uniformly $(0,1)$ distributed random numbers are generated, one for each unique value of the first variable, and treated as probabilities. The probabilities are assigned to the values of the selected variable. Then the probability of obtaining a missing value in that variable is calculated according to the assigned probability and the absolute frequencies of the distinct values in the converted variable. Lastly, indices are randomly selected based on the computed probabilities and the corresponding values are set to missing for each variable. \cite{Thurow2021} argue that this leads to the desired overall missingness rate.
The resulting simulation has $2 \times 2 \times 2 \times 3 = 24$ conditions.
Each combination of the above factors is replicated 1,000 times. 

\subsubsection{Choice of Hyperparameters}
We go with the default choices of hyperparameters if not stated elsewise
\citep{mice,missRanger,mixgb}. For \texttt{mice}, the number of multiple imputations is thus set to \texttt{m=5}, and the number of iterations is set to \texttt{maxit=10}.
Moreover PMM is used with 5 donors.
For the Random Forest and the \texttt{ranger} package, there are more 
hyperparameters that the user has to choose. We use fully grown trees corresponding to \texttt{max.depth=NULL} in the \texttt{ranger} function. For \texttt{mtry} and \texttt{sample.fraction} we also stay at the default value, which are the rounded root of the number of features
and
0.632 times the sample size, respectively. Since the Random Forest is 
computationally expensive we lower the number of trees to speed up the simulation and set
\texttt{num.trees=300} instead of the default value of 500, because the expected performance loss is negligible \citep{probst2018tune}. For the predictive mean matching step
the number of candidates to sample from is set to 5.
For the mixgb imputation method, we also use mostly default hyperparameter settings with two exceptions: The number of multiple imputations is set to \texttt{m=5} for a better comparability with the other imputation approaches. To balance computational efficiency with sufficient convergence, we limit the number of iterations for the imputation process to \texttt{maxit=5}. These values reflect a compromise between computational cost and the stability of imputations needed for robust estimates.

\section{Simulation Study Results}

In this simulation study, various imputation methods were evaluated and compared to address missing data in multilevel designs. The primary objective was to assess the performance of these methods with respect to subsequent inference 
across different settings. To this end, the methods were evaluated using rejection rates and coefficient bias. Rejection rates assess 
Type-I error control (in case of a true null hypothesis) and statistical power (in case of a false null hypothesis), while coefficient bias provides insight into the precision and reliability of the estimated parameters. Evaluating different settings helps to assess the overall robustness of the imputation methods.

Since the overall conclusion from the results of the random intercept model aligns with that of the random slope model, we focus on discussing the former. The results for the random slope model are provided in the supplement.

\subsection{Rejection rates}

\begin{figure}[h]
    \centering
    \begin{subfigure}[b]{1\textwidth}
        \centering
        \includegraphics[width=1\linewidth]{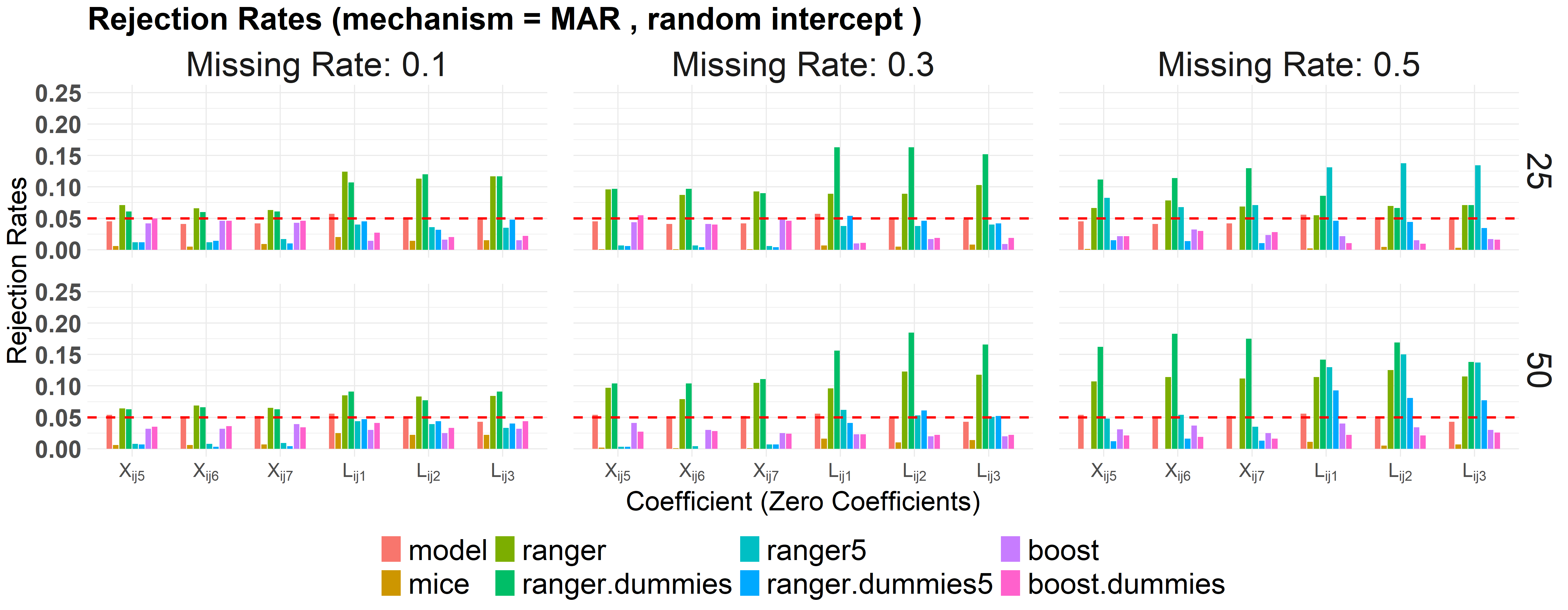}
         \caption{MAR results.}\label{subfig:MAR1}
    \end{subfigure}

    \vspace{0.5cm} 

    \begin{subfigure}[b]{1\textwidth}
        \centering
        \includegraphics[width=1\linewidth]{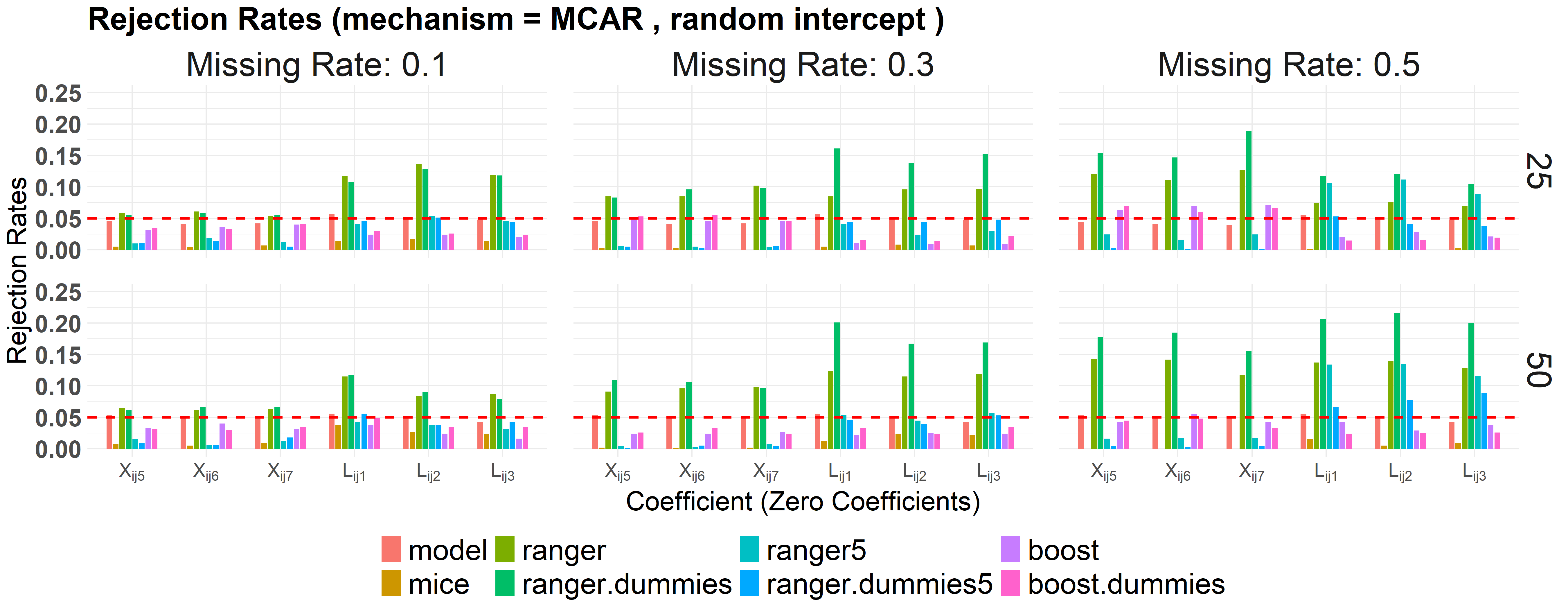}
         \caption{MCAR results.}\label{subfig:MCAR1}
    \end{subfigure}

    \caption{Estimated Type I errors for random intercept designs. The red dotted line represents the 5\% significance level.}
    \label{fig:combined}
\end{figure}

\textit{Type-I Error.} Figure \ref{fig:combined} shows the average rejection rates of $H_0$ across replications for random intercept models, categorized by the missingness mechanism.  
Here, $H_0$ states that the variable (for Level 1) or the cluster variable (for Level 2) has no effect on the dependent variable.
Only coefficients that are truly zero ($X_{ij4},X_{ij5},X_{ij6},L_{ij4},L_{ij5},L_{ij6}$) are included in the figure, thus reflecting the estimated Type I errors. As seen in Subfigures \ref{subfig:MAR1} and \ref{subfig:MCAR1}, $\texttt{missranger}$ without predictive mean matching (displayed in the figures as $\texttt{ranger}$) exceeds the significance level (with and without dummies) for every  coefficient across all combinations of missingness mechanism, missing rate, and cluster size. In some cases, the rejection rates are larger than 20\%, despite the significance level being set to 5\%, highlighting the importance of PMM. Consequently, we exclude Random Forests without PMM from the subsequent power analysis.
For a cluster size of 25 and 50\% missingness, the Type I error with \texttt{missranger} and predictive mean matching (displayed as \texttt{ranger5}) is larger than desired for Level 2 features, though the test maintains the significance level when using dummies. This suggests that dummies can be crucial. However, for a cluster size of 50 and 50\% missingness, all Random Forest approaches result in  estimated Type I errors exceeding 0.05 for the Level 2 features. This holds for both, MCAR and MAR. Notably, for the Level 1 variables without an effect ($X_{ij4}, X_{ij5}, X_{ij6}$), the null hypothesis ($H_0$) was rejected in almost none of the cases for MICE and \texttt{missranger} with PMM, i.e.
the Type I error rate is close to zero, indicating that these methods are overly conservative.

As shown in Figure \ref{fig:combined}, among the different imputation methods evaluated, only MICE and $\texttt{mixgb}$ (both with and without dummies) consistently achieved a Type I error rate below 5\%.\\

\textit{Power.} Figure \ref{fig:combined} displays the rejection rates for the non-zero coefficients, i.e. the estimated power. While the plots for MAR and MCAR are very similar, the missingness rate clearly plays an important role: For 10\% missingness, the Level 1 effects are almost always detected with only minor differences between the imputation methods. However, for the Level 2 variables, MICE outperformed the other methods, with a power approximately 0.3 higher than the others for a cluster size of 25. For 30\% missingness, MICE remains the benchmark for detecting effects on Level 2, but for the Level 1 variables differences between the imputation methods start to occur: 
The rejection rate for the second coefficient for $\texttt{mixgb}$ and $\texttt{mixgb}$ with dummies is better than the rejection rate for MICE, where the variants with dummies seem to have a slight advantage. This holds independent of the cluster size. 

\begin{figure}[h]
    \centering
    \begin{subfigure}[b]{1\textwidth}
        \centering
        \includegraphics[width=1\linewidth]{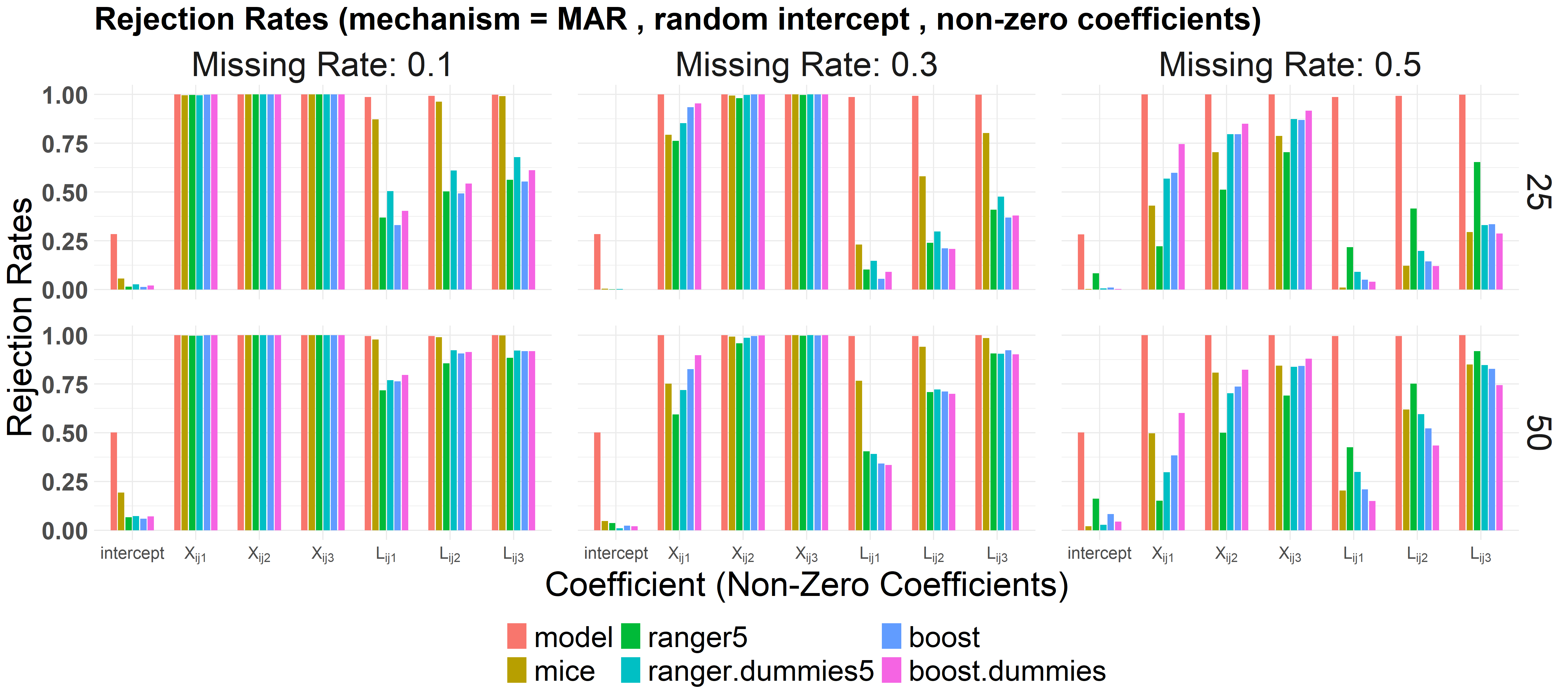}
         \caption{MAR results.1}\label{subfig:MAR2}
    \end{subfigure}

    \vspace{0.5cm} 

    \begin{subfigure}[b]{1\textwidth}
        \centering
        \includegraphics[width=1\linewidth]{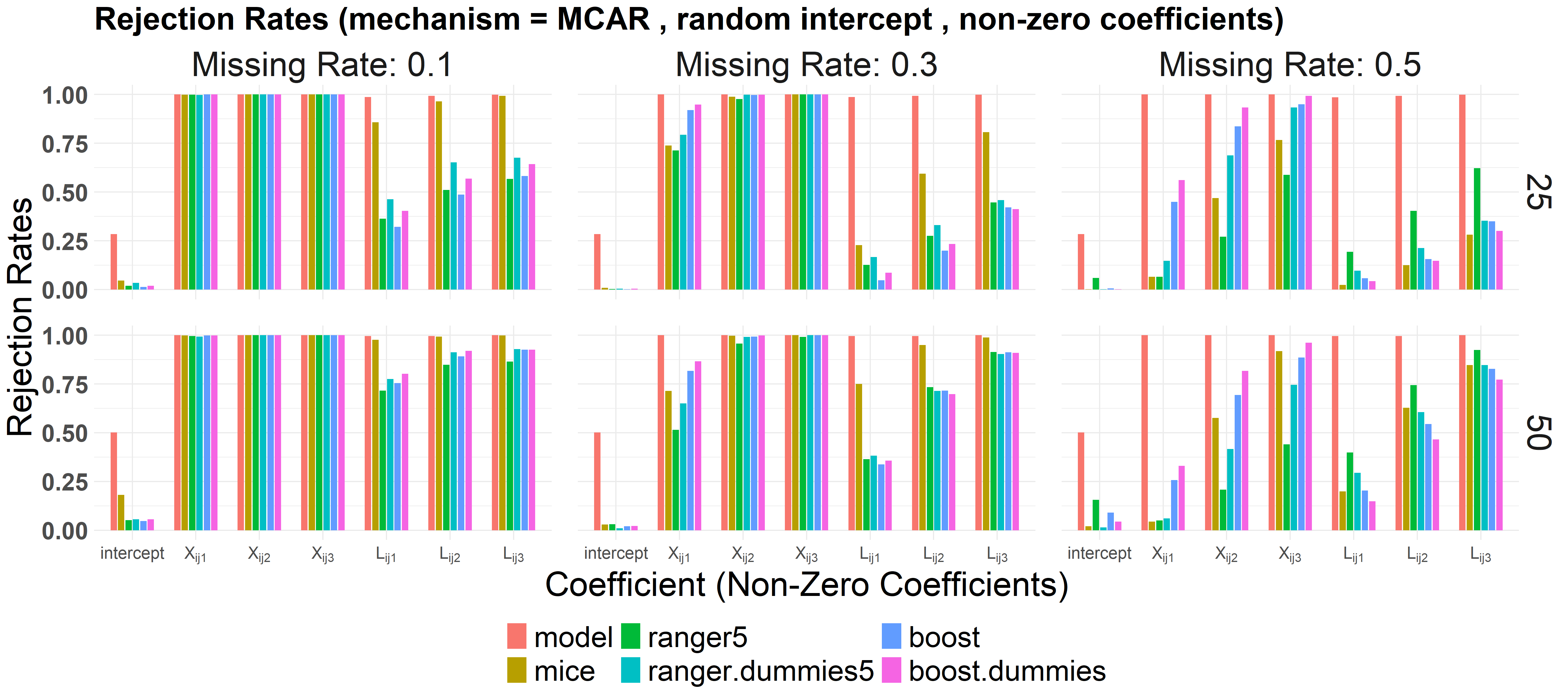}
         \caption{MCAR results.2}\label{subfig:MCAR2}
    \end{subfigure}

    \caption{Power for random intercept designs}
    \label{fig:combined2}
\end{figure}

For the rejection rate of the second coefficient, which has a value of 0.5, $\texttt{mixgb}$ and $\texttt{mixgb}$ with dummies outperform MICE, with the dummy-enhanced version showing a slight advantage. This remains consistent across all cluster sizes. The variants of \texttt{missranger} performed worse than MICE and the $\texttt{mixgb}$ methods. At 50\% missingness, the advantage of $\texttt{mixgb}$ with dummies for Level 1 variables becomes more pronounced. For each cluster size and missingness mechanism, $\texttt{mixgb}$  with dummies sets the benchmark. Particularly for MCAR, $\texttt{mixgb}$ with dummies significantly outperforms MICE. For Level 2 variables, $\texttt{mixgb}$ with dummies performs similarly to MICE at a cluster size of 25, whereas at a cluster size of 50, MICE has a slight advantage. This pattern is consistent for both missingness mechanisms. Therefore, at 50\% missingness, $\texttt{mixgb}$ with dummies emerges as the preferred method, even though MICE benefits from knowing the linear structure of the model equation.

\subsection{Coefficient bias}

\begin{figure}[h!]
    \centering
    \includegraphics[width=1\linewidth]{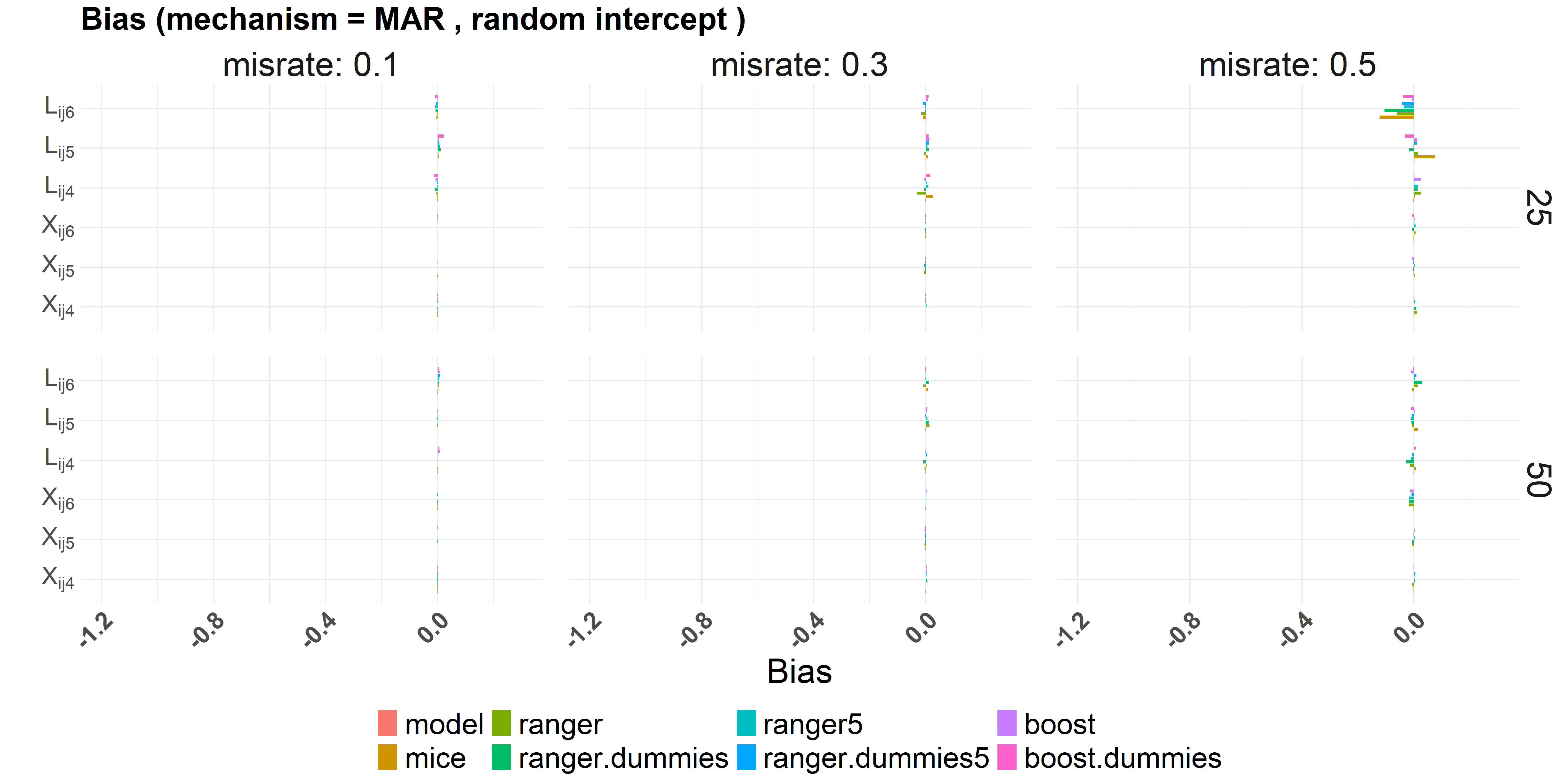}
    \vspace{0.5cm} 
    \includegraphics[width=1\linewidth]{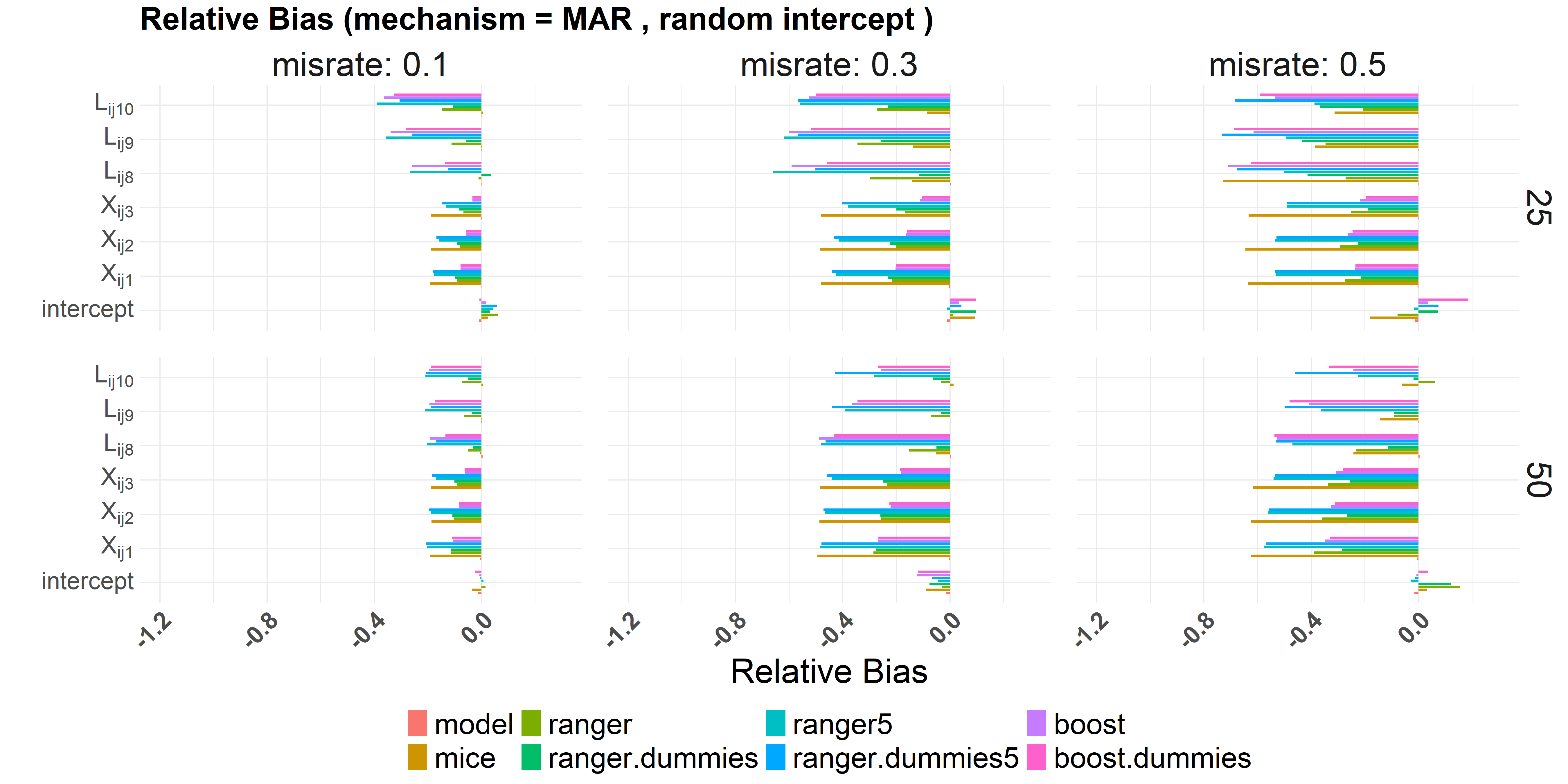}
    \caption{Bias for the true zero coefficients and Relative Bias for the non-zero coefficients}
    \label{fig:combined_bias}
\end{figure}

Figure \ref{fig:combined_bias} shows the coefficient bias for each method under random intercept models with MAR. For the true zero coefficients at both Level 1 and Level 2, we observe that the bias values are consistently close to zero. Even with increasing missingness or decreasing cluster size, no significant deviations from zero are found.
A more interesting observation is the relative bias (adjusted for effect size) for the non-zero coefficients.
The relative bias is used to normalizes the bias by the true parameter value, making it easier to interpret and compare across different coefficient-sizes.
It appears that all imputation methods induce a negative bias, with the exception of the intercept, where a positive bias occurs in some cases. However, the relative bias for the intercept is negligible compared to that of the slope coefficients. These findings regarding relative bias align closely with the rejection rates observed earlier.
\texttt{mixgb}, both with and without dummies, exhibits a relatively high relative bias for the Level 2 variables. In contrast, MICE produces the lowest bias among the tested imputation methods for Level 2 variables, regardless of the missingness rate. On the other hand, for Level 1 variables, MICE shows a larger bias than the boosting methods, and this difference increases with increasing missing rate. The Random Forest methods also show a relatively large bias for both Level 1 and Level 2 variables.

\section{Discussion}

This paper critically evaluates the performance of novel tree-based imputation methods for handling missing data in hierarchical data structures, focusing on  Random Forest (\texttt{missranger}) and XGBoost (\texttt{mixgb}) based approaches with and without dummy modeling.   Through a comprehensive simulation study, we contrast these novel techniques with the state-of-the art MICE approach, focusing in particular on bias and inference. The evaluation considers Level 1 and Level 2 variables across two different data generation processes (random intercept and random slope models), three missingness rates (10\%, 30\% and 50\%), and two missingness mechanisms (MCAR and MAR). Our results indicate that MICE is characterized by consistent accuracy in rejection rates when using the \verb|2l.norm| method for Level 1 variables and \verb|2lonly.pmm| for Level 2 variables. This consistency underscores MICE's robustness to hierarchical data and closely approximates true rejection rates (i.e., type I error and power).
Our results are consistent with the findings in \cite{grundluedtkeLEVEL2}, demonstrating the practical usefulness of MICE while dealing with missingness in multilevel structures, especially in the case of missingness at Level 2. Similar to \cite{endetal16}, we also found that MICE was superior not only for random slope but for random intercept models, provided that the imputation model was specified according to the data generation process and the missing rate was low (10\%).

For Level 1 coefficients, 
boosting (\texttt{mixgb}) with dummies consistently outperforms other methods, including MICE, across all missingness mechanisms when the missingness rate is 30\% or 50\%. At 30\% missingness, however, MICE has more power for Level 2 variables, making the choice of imputation method dependent on the practitioner’s priorities and the relative importance of Level 1 versus Level 2 variables in the specific application.
At 50\% \texttt{mixgb} (with and without dummies) proved to be the preferable alternative in comparison to MICE regardless of missing mechanism or cluster size.

 The analysis of (relative) coefficient bias corroborates the findings from the rejection rates, revealing a larger relative bias for MICE in Level 1 variables and a greater relative bias for the boosting methods in Level 2 variables. 
 
The \texttt{missranger} methods, on the other hand, fell slightly behind compared to the boosting methods and MICE for almost settings and cannot be recommended as an imputation method in our multilevel settings. In particular, \texttt{missranger} without PMM showed highly increased type I errors.

While the rejection rate differences between imputation methods slightly vary across the random intercept  and random slope models, the overall trends remain consistent regardless of the data-generating process.
The results do not necessarily indicate that tree-based methods with dummies (also including dummies for the Level 2 cluster) outperform their standard counterpart without dummies. For example, under MCAR and MAR, standard 
 \texttt{mixgb} showed slightly higher power than 
 \texttt{mixgb}  with dummies for Level 2 coefficients when the missingness rate was 50\% in both random intercept and random slope.  Meanwhile, adjusted \texttt{mixgb} sometimes exhibited higher power for Level 1 coefficients. In some 
situations (e.g. 50\% missingness and cluster size 25) the introduction of dummies helped to decrease the type I error to the desired level. But for most scenarios the introduction of dummies did not alter the results much.

\subsection{Limitations, strengths and outlook}

Although the study is comprehensive, it is not without limitations. The simplicity of the linear data structure used in our simulations, characterized by a low number of variables, contrasts with the often complex and variable-rich data sets encountered in real-world scenarios. As a result, the generalizability of our results to more complicated data sets is uncertain. 
The linear data generation processes could also be a reason why MICE has more reliable rejection rates than the tree-based methods in case of low missing rates. Typical advantages of tree-based methods 
include their ability to handle non-linear effects and a large number of variables. However, 
these strengths cannot shine  in our relatively low-dimensional data typical for applications of LMMs. In addition, among all imputation methods, MICE best aligns with the true underlying data generation process. However, this process is generally not known in empirical data.
The fact that \texttt{mixgb} was still able to outperform MICE for increasing missing rates underlines the potential of tree-based imputations methods for multilevel data.

Future research investigating the impact of non-linear variable effects and a larger number of variables on both rejection rate and bias could provide a more complete understanding of the methods' performance n more complex data settings.  In particular, simulating data beyond a linear mixed model could be insightful, such as using a semi- or non-parametric data generating model that remains plausible enough for an LMM-analysis.

The strength of the current study, however, lies in its novelty. We introduce and evaluate the combination of the latest tree-based imputation methods and a simple multilevel adjustment via dummy variables against established techniques such as MICE. This approach provides new insights into the evolving landscape of data imputation methods, especially in the context of hierarchical data structures.

Another interesting area to explore would be to improve the performance of tree-based methods. This could involve experimenting with multivariate tree-based methods  \citep{SegaX2011,schmid2023tree}. Tree-based methods with random effects have been developed \citep{bergonzoli2024ordinal,buczak2024,FokkemaEA2018,
FokkemaEW2020,
HajjemBL2014,
HajjemLB2017,
SelaS2012}, but still have to be adapted to missing data. Another possible approach could be to first average all data at level 2 and impute missing data at this aggregate level, and then merge only the level 2 variables with the individual level dataset.

In conclusion, our study underscores the continued effectiveness of MICE in dealing with hierarchical data, particularly in terms of rejection rates. However, the emerging tree-based methods, especially \texttt{mixgb}, show potential for larger missing rates, suggesting their usefulness as alternatives in certain contexts. This dual finding opens new avenues for future research and practical applications in data imputation, highlighting the dynamic nature of the field.

\section*{Acknowledgement}

The authors gratefully acknowledge the computing time provided on the Linux HPC cluster at Technical University Dortmund (LiDO3), partially funded in the course of the Large-Scale Equipment Initiative by the Deutsche Forschungsgemeinschaft (DFG, German Research Foundation) as project 271512359.

\section*{Funding}

The project ``From Prediction to Agile Interventions in the Social Sciences (FAIR)'' is receiving funding from the programme ``Profilbildung 2020'', an initiative of the Ministry of Culture and Science of the State of Northrhine Westphalia. The sole responsibility for the content of this publication lies with the authors.

Nico Föge was supported by the Deutsche Forschungsgemeinschaft (DFG, German Research Foundation) - 314838170, GRK 2297 MathCoRe. 

\section*{Contributions}

Author contributions: \insertcreditsstatement

\bibliographystyle{apacite}
\bibliography{sample}

\newpage
\appendix 
\section{Random slope results}

\subsection{Rejections rates}

\begin{figure}[h]
    \centering
    \begin{subfigure}[b]{1\textwidth}
        \centering
        \includegraphics[width=1\linewidth]{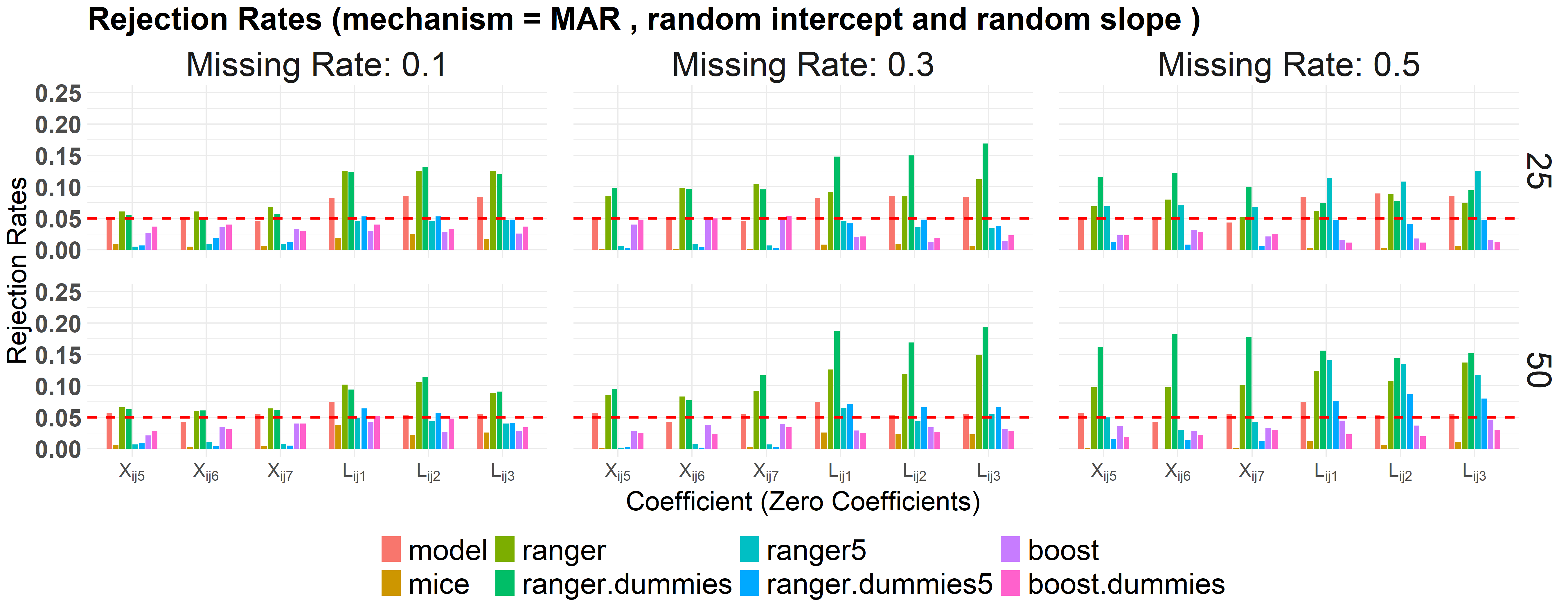}
         \caption{MAR results.}\label{subfig:MAR3}
    \end{subfigure}

    \vspace{0.5cm} 

    \begin{subfigure}[b]{1\textwidth}
        \centering
        \includegraphics[width=1\linewidth]{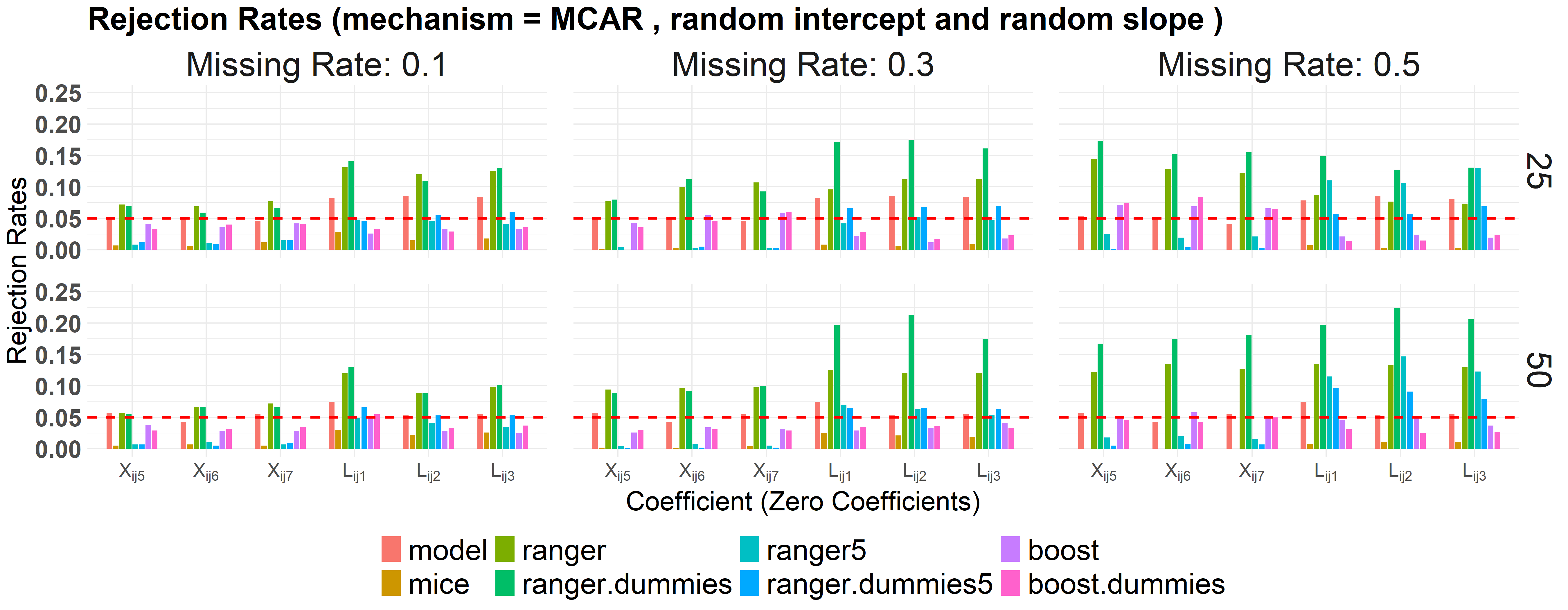}
         \caption{MCAR results.}\label{subfig:MCAR3}
    \end{subfigure}

    \caption{Type-I-Error for random intercept and random slope designs}
    \label{fig:combined3}
\end{figure}

For both random intercept and random slope models, Figure \ref{fig:combined3} presents the Type I error rates across replications, grouped by the missingness mechanism. The results are broadly consistent with those observed for the random intercept model. Notably, the Random Forest imputation method without PMM fails to maintain the significance level across all variables, regardless of the combination of missingness rate, missingness mechanism, or cluster size. This issue is particularly pronounced for Level 2 variables, where Type I error rates are significantly higher than 0.05. For the MCAR mechanism, Type I errors exceed 0.20 in some cases. Consequently, we exclude Random Forest without PMM from the power analysis.

For scenarios with 50\% missingness, even the Random Forest method with PMM exceeds the acceptable significance level. While MAR with a cluster size of 25 shows that the introduction of PMM can reduce Type I error rates below 0.05, in other scenarios, the Random Forest with PMM and dummy variables still produces Type I error rates above the acceptable threshold. In contrast, the boosting methods and MICE consistently maintain the significance level, demonstrating their robustness in this context.

\begin{figure}[h]
    \centering
    \begin{subfigure}[b]{1\textwidth}
        \centering
        \includegraphics[width=1\linewidth]{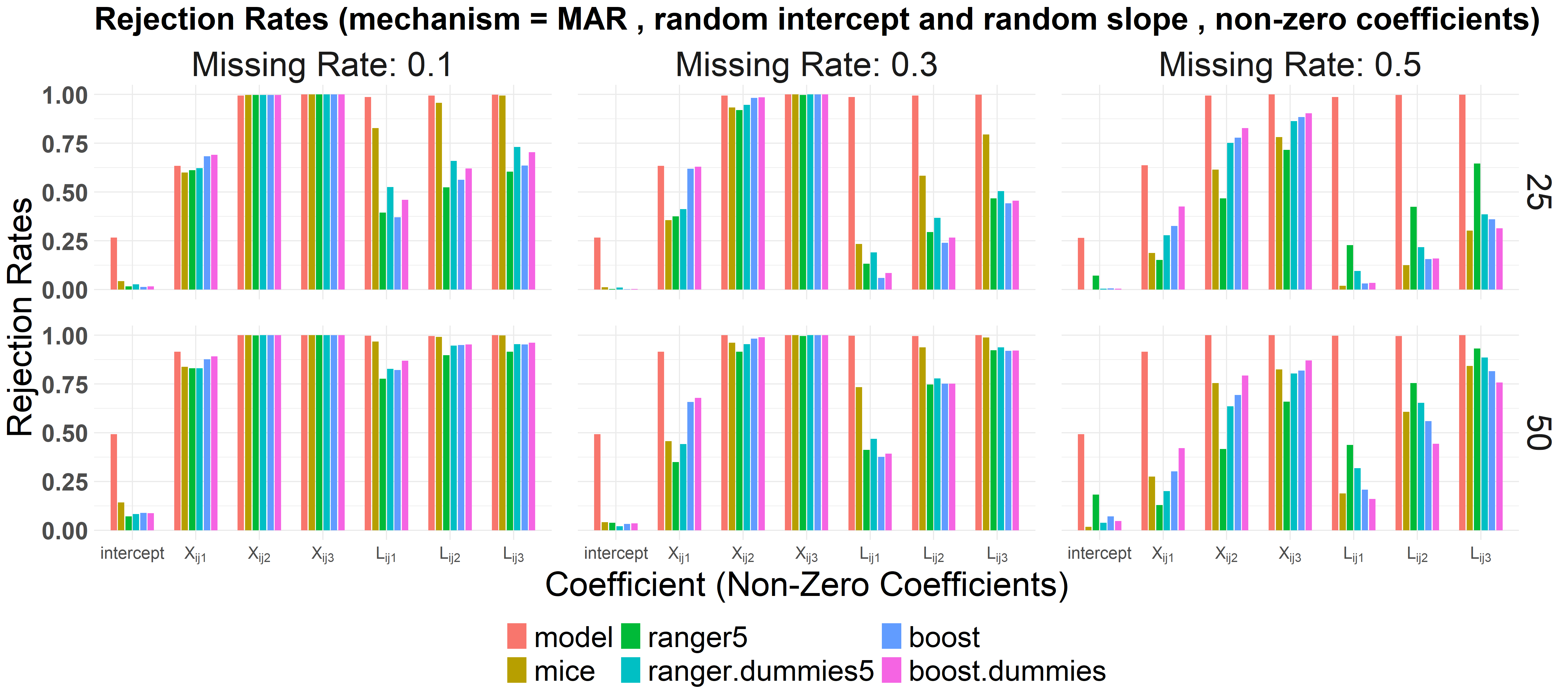}
         \caption{MAR results.}\label{subfig:MAR4}
    \end{subfigure}

    \vspace{0.5cm} 

    \begin{subfigure}[b]{1\textwidth}
        \centering
        \includegraphics[width=1\linewidth]{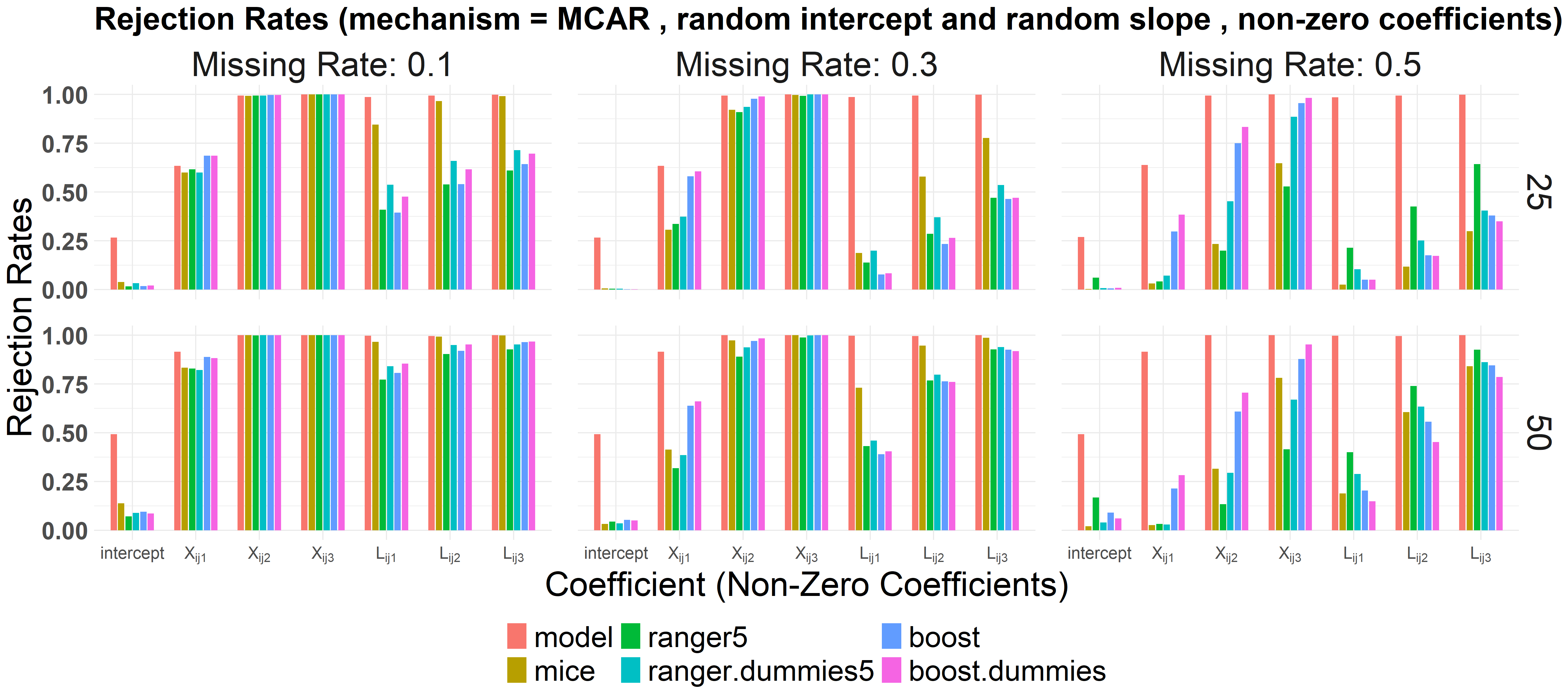}
         \caption{MCAR results.}\label{subfig:MCAR4}
    \end{subfigure}

    \caption{Power for random intercept and random slope designs}
    \label{fig:combined4}
\end{figure}

In Figure \ref{fig:combined4}, the rejection rates for the random intercept and random slope models are presented. The primary differences in the comparison of the imputation methods arise again with varying missing rates.
For 10\% missingness, all methods exhibit similar rejection rates for the Level 1 variables. However, for the Level 2 variables, MICE clearly outperforms the other imputation methods, with its advantage being more pronounced when the cluster size is 25. For a cluster size of 50, MICE remains the best-performing method, but the tree-based methods show performance closer to that of MICE.\\
Similar to the random intercept model, a missingness rate of 30\% also leads to mixed results for the random intercept and random slope models. For the Level 1 variables, \texttt{mixgb} (with and without dummies) outperforms all other imputation methods, with a slight advantage for the version using dummies. In contrast, for the Level 2 variables, MICE clearly emerges as the best option, with only minor differences observed among the tree-based methods.\\
At 50\% missingness, the advantage of \texttt{mixgb} (with and without dummies) for Level 1 variables increases significantly under MCAR. For MAR, the boosting method with dummies remains the top choice for Level 1 variables, though the performance gap to MICE is smaller. The largest rejection rates for Level 2 variables at 50\% missingness are observed with the Random Forest method without dummies; however, these results are unreliable as this method leads to an inflated type I error rate.

Regardless of the missing data mechanism or cluster size, \texttt{mixgb} with dummies generally performs slightly better than \texttt{mixgb} without dummies. Overall, the Level 2 rejection rates for MICE are comparable to those of the boosting methods. However, given the advantages for the Level 1 variables, \texttt{mixgb} with dummies should be preferred over MICE at a 50\% missingness rate.

\subsection{Coefficient bias}

\begin{figure}[h!]
    \centering
    \includegraphics[width=1\linewidth]{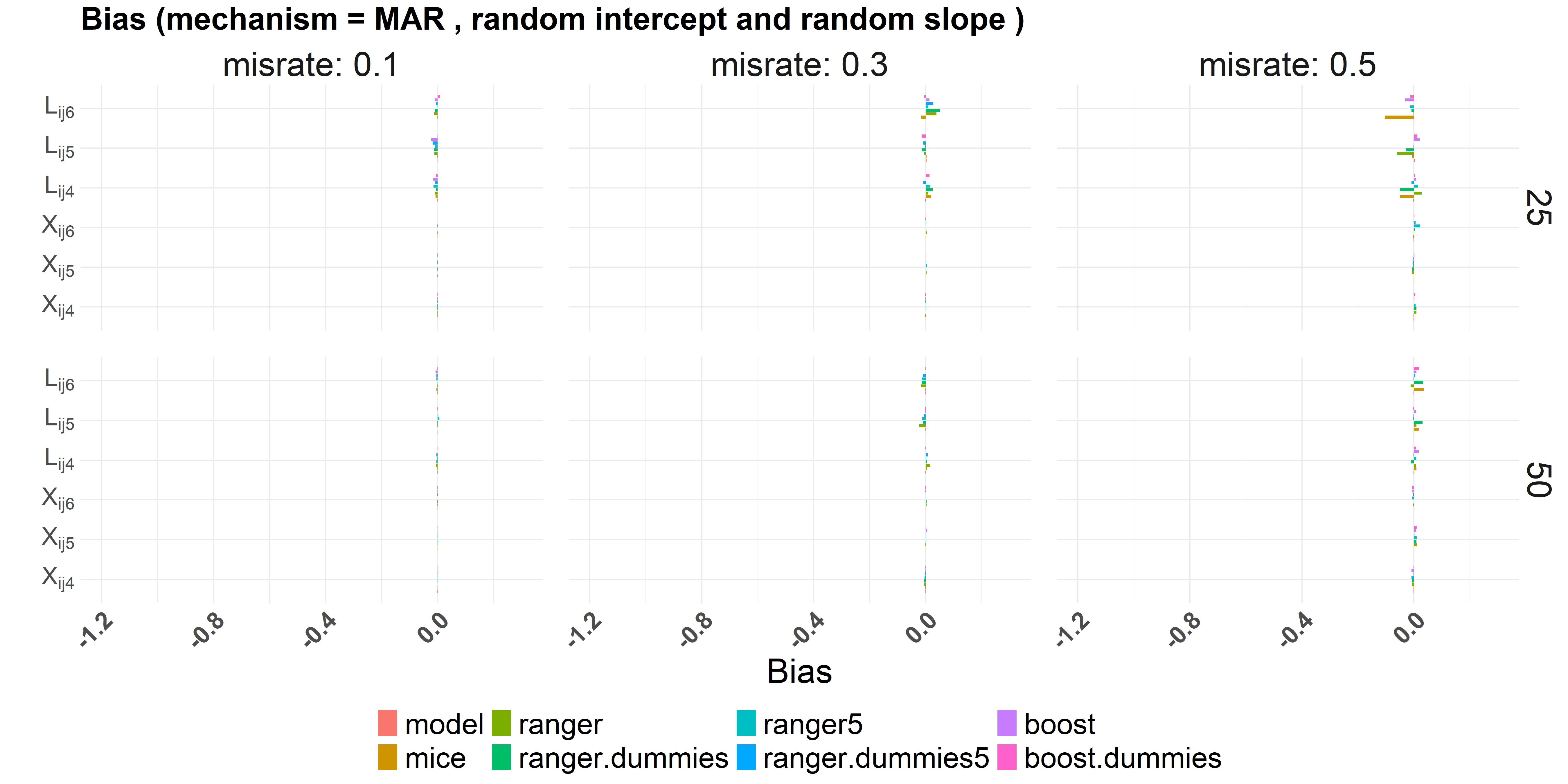}
    \vspace{0.5cm} 
    \includegraphics[width=1\linewidth]{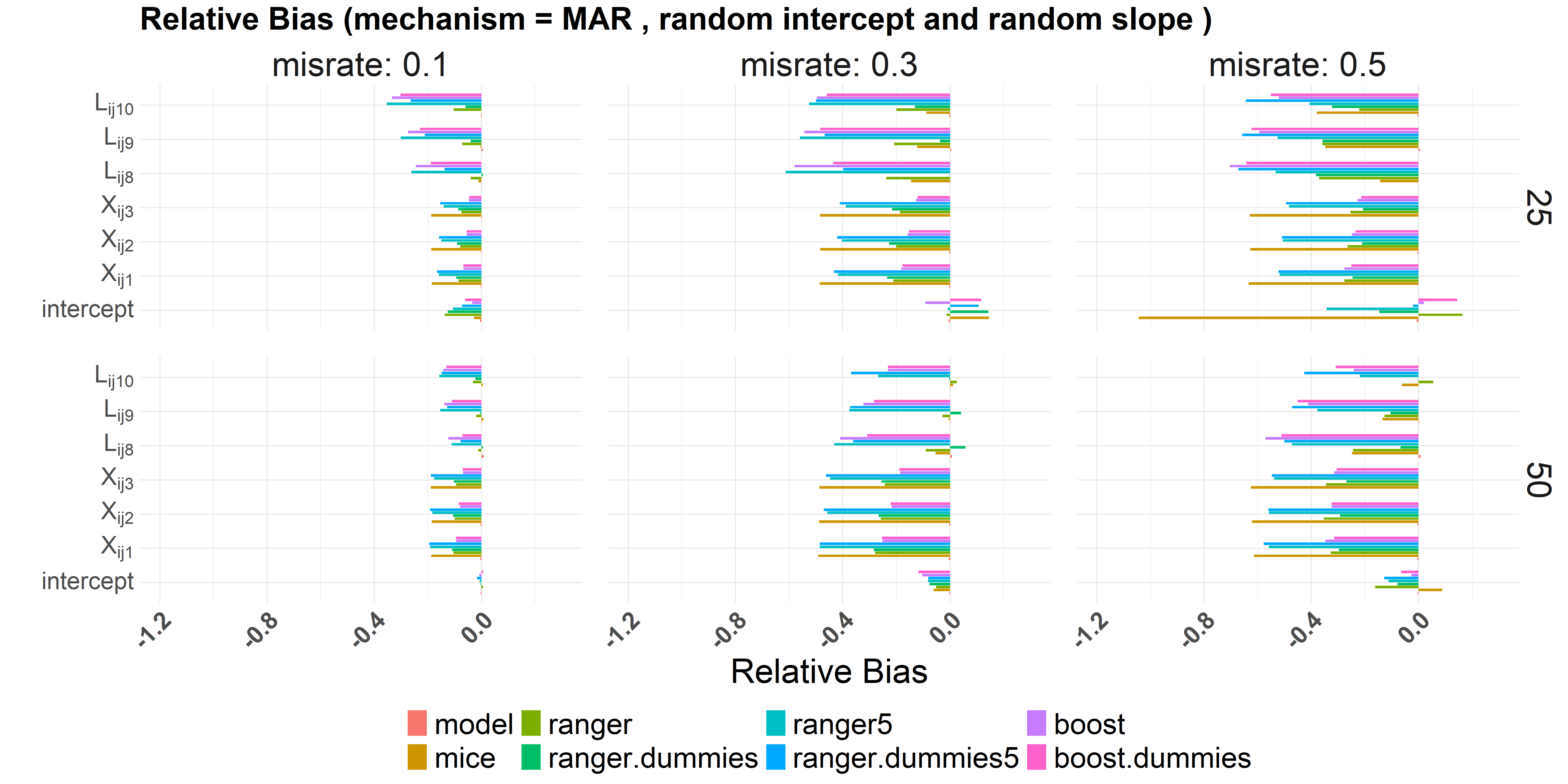}
    \caption{Bias for the true zero coefficients and Relative Bias for the non-zero coefficients}
    \label{fig:combined_bias2}
\end{figure}

Figure \ref{fig:combined_bias2} presents the coefficient bias and relative bias for each method under the random intercept and random slope model. The first plot shows the bias for the true zero coefficients at Level 1 and Level 2, while the second plot shows the relative bias (adjusted for effect size) for the non-zero coefficients. Overall, the pattern is quite similar to the previous figure, with a few notable differences.

For the true zero coefficients at both levels, the bias values remain close to zero, even with increasing missingness rates or decreasing cluster sizes. This suggests that the imputation methods do not introduce any significant bias for coefficients that are expected to be zero.

The more interesting observation concerns the relative bias for the non-zero coefficients. Similar to the previous figure, all imputation methods induce a negative bias, but the magnitude of this bias varies across methods and scenarios. Notably, MICE shows a relatively large negative bias for the intercept when the cluster size is 25 and the missingness rate is 50\%. This stands in contrast to the other methods, which do not show a similarly pronounced bias for the intercept. 

The relative bias for the slope coefficients aligns with the findings from the previous analysis. \texttt{mixgb} (both with and without boosting) exhibits a relatively high bias for Level 2 variables, especially when compared to MICE, which shows the lowest bias for the Level 2 coefficients, regardless of missingness rate or cluster size. For Level 1 variables, MICE shows a larger bias than the boosting methods, and this difference increases as the missingness rate rises.

In general, the Random Forest methods show a relatively large bias for both Level 1 and Level 2 variables across the different missingness rates and cluster sizes. For the MCAR missingness mechanism, the results remain largely consistent, with only minor differences compared to the MAR scenario. The corresponding plots for MCAR are included in the appendix.

\clearpage

\end{document}